\documentstyle[preprint,aps]{revtex}
\begin{document}
\draft
\preprint{IASSNS-HEP-94/27}
\title{Exact Dynamical Correlation Functions of Calogero-Sutherland \\
Model and One-Dimensional Fractional Statistics}
\author{Z. N. C. Ha}
\address{
School of Natural Sciences, Institute for Advanced Study, \\
Princeton, New Jersey 08540
}
\date{April 18, 1994}
\maketitle
\begin{abstract}
One-dimensional (1D) model of non-relativistic particles
with inverse-square interaction potential known
as Calogero-Sutherland Model (CSM) is shown to possess
fractional statistics.  Using the theory of Jack
symmetric polynomial the exact dynamical density-density
correlation function and the one-particle Green's function
(hole propagator) at any rational interaction coupling
constant $\lambda = p/q$ are obtained and used to show clear evidences
of the fractional statistics.
Motifs representing the eigenstates of the model are
also constructed and used to reveal the fractional {\it exclusion}
statistics (in the sense of Haldane's ``Generalized Pauli Exclusion
Principle'').
This model is also endowed with a natural {\it exchange }
statistics (1D analog of 2D braiding statistics)
compatible with the {\it exclusion} statistics.
\newline{(Submitted to PRL on April 18, 1994)}
\end{abstract}
\narrowtext

The quantum particles obeying fractional statistics known as anyons
have been a subject of intense study since the discovery of
the fractional quantum Hall effect and the High $T_c$ super-conductors
\cite{book1}.  The subject, however, is far from complete.
In fact, even the simplest model of anyons, namely the ideal gas,
has not yet been fully understood. To add to the confusion,
there are now two seemingly non-equivalent definitions of fractional
statistics.  While the popular definition of anyons is based on the
quantum phase arising from the exchange of particles \cite{book1},
Haldane's definition \cite{duncan1} is based on so called the
``Generalized Pauli Exclusion Principle.''
The main difference in the two approach
is that while in the former the statistics
is assigned to the Newtonian point particles,
in the latter it is obeyed
by the elementary excitations of condensed matter system.

There are no fully solvable two-dimensional models where the ideas of
fractional statistics can be rigorously tested.  In one-dimension,
however, there is such a model known
as Calogero-Sutherland Model (CSM) \cite{csm}.
I show in this Letter that the two definitions of fractional statistics
can coexist in CSM without any inconsistency.
Following Ref.\ \cite{hahal},  I construct the motifs
for all the excited states and explicitly demonstrate
that the quasi-particles and quasi-holes obey the
Haldane's exclusion statistics.
I also solve the exact ground state dynamical
density-density correlation function (DDDCF)
and the hole propagator part of the one-particle Green's function (HPOGF)
and show that the contributing intermediate states
involve only a finite number of quasi-particles (holes) consistent with the
ideal ``anyon'' gas structure of this model.
In calculating the correlation function and
propagator a new mathematical technique based on the
theory of Jack symmetric orthogonal polynomial \cite{stanly} is used.

The CSM Hamiltonian, which describes a system of $N$
non-relativistic particles interacting with inverse-square exchange,
is given by
\begin{equation}
H = -\sum_{j=1}^N {\partial^2 \over \partial x_j^2} +
\sum_{j<l}{2\lambda(\lambda - 1) \over d^2(x_j - x_l)},
\label{hamil}
\end{equation}
where $\hbar^2/2m = 1$ and
$d(x_i-y_j)$ is the chord distance between the $i$th and
$j$th particles on a ring of length $L$, and is equal to
$|(L/\pi)\sin(\pi(x_i-y_j)/L)|$. The dimension-less interaction coupling
constant $\lambda$ is a positive real number that specifies the natural
statistics of this model.   For the special
values of $\lambda = 1/2$, $1$, and $2$, the model is related
to the orthogonal, unitary and simplectic random matrix theory \cite{dyson},
respectively, and DDDCF for those values have previously been found by the
supersymmetry technique \cite{simons}.
For $\lambda = 2$, the full one-particle Green's function has also
been found \cite{haldzern,zernhald}.

One of the main results of this Letter is the calculation of the dynamical
density-density correlation function
at any rational interaction parameter $\lambda = p/q$. It is given by
\begin{equation}
\langle0|\rho(x,t)\rho(0,0)|0\rangle = C\ Re\prod_{i=1}^q
\left(\int_0^\infty dx_i \right)
\prod_{j=1}^p \left(\int_0^1 dy_j\right) Q^2 F(q,p,\lambda|\{x_i,y_j\})
e^{i(Q x-E t)},
\label{ddc}
\end{equation}
where $Q$ (momentum) and $E$ (energy) are given by
\begin{eqnarray}
Q & = & 2\pi\rho\left(\sum_{j=1}^q x_j + \sum_{j=1}^p y_j\right) \\
E & = & (2\pi\rho)^2\left(\sum_{j=1}^q \epsilon_P(x_j) +
\sum_{j=1}^p \epsilon_H(y_j)\right),
\end{eqnarray}
where $\rho = N/L$, $\epsilon_P(x) = x(x+\lambda)$ and
$\epsilon_H(y) = \lambda y(1-y)$. $x_j(\epsilon_P)$ and $y_j(\epsilon_H)$
are normalized momentum (energy) of
the quasi-particles and the quasi-holes, respectively.
The normalization constant $C$ is given by
\begin{eqnarray}
A(m,n,\lambda) & = & {\Gamma^m(\lambda) \Gamma^n(1/\lambda)
\over \prod_{i=1}^m \Gamma^2(p-\lambda(i-1))
\prod_{j=1}^n \Gamma^2(q-(j-1)/\lambda)} \prod_{j=1}^n
{\left(\Gamma(q-(j-1)/\lambda) \over \Gamma(1-(j-1)/\lambda)\right)}^2, \\
C & = & {\lambda^{2p(q-1)} \Gamma^2(p)\over 2\pi^2 p!q!} A(q,p,\lambda).
\end{eqnarray}
Finally, the form factor $F(q,p,\lambda|\{x_i,y_j\})$ is given by
\begin{equation}
F(q,p,\lambda|\{x_i,y_j\}) =
\prod_{i=1}^q \prod_{j=1}^p (x_i + \lambda y_j)^{-2}
{\left(\prod_{i<j} (x_i - x_j)^2\right)^\lambda
\left(\prod_{i<j} (y_i - y_j)^2\right)^{1/\lambda}
\over
\prod_{i=1}^q \epsilon_P(x_i)^{1-\lambda}
\prod_{j=1}^p \epsilon_H(y_j)^{1-1/\lambda}}.
\label{formfac}
\end{equation}
The results of Simons, et al. at $\lambda = 1/2$, $1$ and $2$
with an appropriate change of variables agree with Eq.\ (\ref{ddc})
up to the normalization constant.  Recently, based on a mapping to
2D QCD Minahan guessed correctly
the form factor for the integer values of $\lambda$ and their
duals $1/\lambda$ \cite{minahan}.  Haldane also guessed the
correct form factor for the rational $\lambda = p/q$ based on
the information given by Simons, et al., Galilean invariance,
and asymptotic agreement with U(1) conformal field theory
\cite{haldane}.

The ground state of SCM is given by
\begin{equation}
\Psi_0 = \prod_{j>l} (z_j-z_l)^\lambda \prod_{j=1}^N z_j^{J_0}
\label{gs}
\end{equation}
where $z_j = \exp(i 2\pi x_j/L)$ and $J_0 = -\lambda(N-1)/2$.
If a general eigenfunction with energy $E$ is written
as $\Psi = \Psi_0 \Phi$, then $\Phi$ is an
eigenstate of the following new effective Hamiltonian $H'$
\begin{equation}
H'\Phi = \sum_i (z_i\partial_{z_i})^2 \Phi +
\lambda \sum_{i<j} {z_i + z_j \over
z_i - z_j}(z_i\partial_{z_i} - z_j\partial_{z_j})\Phi = \varepsilon \Phi,
\label{h2}
\end{equation}
where $\varepsilon = (L/2\pi)^2(E-E_0)$.  An amazing coincidence
happens if $\Delta = (\lambda N - \lambda - 1)\sum_j z_j\partial_{z_j}$
is added to Eq.\ (\ref{h2}).  A complete set of
linearly independent solution of the resulting equation
is known in mathematical
literature as the Jack polynomial
$J^{1/\lambda}_\kappa(z_1,\ldots,z_N)$ \cite{stanly}.  The index
$\kappa = (\kappa_1,\kappa_2,\ldots,\kappa_N)$
is partition of non-negative integers and is essentially a set of
bosonic quantum numbers
used to label all the eigenstates of SCM up to global Galilean boosts.
In particular, the ground state is given by the partition of zero
(i.e. $\kappa_1 = \kappa_2 = \ldots = \kappa_N = 0$).
The parts $\kappa_j$'s of partition $\kappa$ are ordered so that
$\kappa_1 \ge \kappa_2 \ge \cdots \ge \kappa_N$.
Since the $J_\kappa$
is a homogeneous symmetric polynomial of degree $|\kappa| = \sum_j \kappa_j$,
it is also an eigenfunction of $\Delta$ with eigenvalue
$(\lambda N - \lambda - 1)|\kappa|$ and thus of $H'$ with
$\varepsilon = \sum_j \kappa_j + \lambda(N+1-2j)\kappa_j$.
If the pseudo-momenta $k_j$'s are defined by
\begin{equation}
Lk_j = 2\pi I_j+ \pi(\lambda-1)\sum_l \mbox{sgn}(k_j-k_l),
\label{pseudomom}
\end{equation}
with $I_j = \kappa_j + (N+1-2j)/2$,
then the eigenenergy and momentum take the following free form:
$E = \frac{\hbar^2}{2m}\sum_j k_j^2$ and $P = \sum_j k_j$.
Using a method developed by Yang and Yang \cite{yang}, Sutherland
calculated the thermodynamics for SCM \cite{suther} and
found that the densities of occupied ($\rho_P$) and
unoccupied ($\rho_H$) $k$ satisfy the following relation:
$\lambda\rho_P(k) + \rho_H(k) = 1$.  This is a statement of {\it broken
particle-hole symmetry} for $\lambda \ne 1$ and an analog of the
Chern-Simons duality. $\rho_P(k)$ satisfies
the following relation \cite{bernardwu},
$(1-\lambda \rho_P)^\lambda (1-(\lambda-1)\rho_P)^{\lambda-1} =
\rho_P \exp((\epsilon(k)-\mu)/T)$ with $\epsilon(k) = k^2$,
which is identical to that of the ideal gas obeying
Haldane's fractional exclusion statistics \cite{wu}.
I emphasize here that
the statistical distribution above is only for the pseudo-particles
and is not related to the statistics of the real particles.

I use three different names for the particles in
the model---real, pseudo-,
and quasi-particles.  The real particles are, of course, the
physical quantum particles described by the canonically conjugate
coordinate and momentum variables $\{x_j, p_j\}$.  The pseudo-particles
are described by the pseudo-momentum operators (see Eq.\ (\ref{pi}))
whose eigenvalues are given by Eq.\ (\ref{pseudomom}).
The quasi-particles are the elementary excitations of the
system. Because the pseudo-particles form an ideal gas,
the quasi-particles are essentially same as the pseudo-particles excited
out of the condensate.  The holes left behind in the pseudo-particle
condensate will be called quasi-holes.  Hence, the name ``pseudo''
and ``quasi'' will be used interchangeably in some cases.

If $\lambda = p/q$, where $p$ and $q$ are relative primes,
the motifs for eigenstates can be constructed.
The {\it ones} and {\it zeroes}
in the motif mean occupied and unoccupied $k_j$, respectively.
{}From Eq.\ (\ref{pseudomom}) the following rules
can be deduced for constructing the motif.
(i) Total of $N$ {\it ones} in the
motif represent the pseudo-particles.  Hence, the charge of
pseudo- and real particle is same.
(ii) {\it The allowed number of
{\bf zeroes} between each pair of {\bf ones} is $p-1+nq$, where $n$ is an
arbitrary non-negative integer.}
Of the $p-1+nq$ allowed zeroes, $p-1$
consecutive {\it zeroes} are {\it bound} to each {\it one} while the rest
of them are {\it un-bound}.
The consecutive $q$ {\it un-bound zeroes} represent a hole.
(iii) A cutoff is introduced to make the
number of {\it un-bound zeroes} finite.
The natural unit for the pseudo-momenta is then $2\pi/qL$ which is an
analog of the flux quantum.
To give some examples,
the ground state $(a)$, allowed $(b)$ and forbidden ($c$) states
for $\lambda = 3/4$ for $N=7$ are represented, respectively, by
\begin{eqnarray*}
&(a)\quad \cdots 00000000100100100100100100100000000\cdots \\
&(b)\quad \cdots 00000000100100100100100100000010000\cdots \\
&(c)\quad \cdots 00000000100100100000100100100100000\cdots
\end{eqnarray*}
The motif (c) violates Rule (ii) and hence not allowed.
The {\it bound zeros} which cause spacings
between the {\it ones} can be viewed as flux attached to the
pseudo-particles.

The motif thus constructed is a powerful tool in visualizing the
fractional statistics.
Destroying a particle (or turning a {\it one} into a {\it zero})
creates $p$ extra {\it un-bound} {\it zeroes}
and in order to have integer number of holes in the condensate
(Rule (ii)),
minimum of $q-1$ extra quasi-particles must be excited out of the
condensate leaving behind total of $qp$ {\it un-bound zeroes} which
break up into $p$ holes.
Hence, $q$ particles leave behind $p$ holes.
This is a generalization of the Pauli exclusion principle which explains
intuitively
why the density-density correlation function has $q$ quasi-particles and
$p$ quasi-holes when $\lambda = p/q$.
It will be shown explicitly that the only states that
contribute to the DDDCF are indeed this minimal excitation.

Before I begin to discuss the method used to calculate the DDDCF,
some notations need to be defined.
First, a diagram ${\cal D}(\kappa)$ is defined to be rows and columns of
boxes labeled by
$\{(i,j):1\le i \le l(\kappa), 1 \le j \le \kappa_i\}$, where
$l(\kappa)$ denotes the number of non-zero $\kappa_j$.  The label
$i$ and $j$ are row and column indices of the diagram with $l(\kappa)$
rows of lengths $\kappa_j$.  Second, the conjugate of $\kappa$ denoted by
$\kappa'$ is obtained from $\kappa$ by changing all the rows to columns
in non-increasing order \cite{macdonald}.
Each row (column) corresponds to the quasi-particle (hole) excitations.
Third, a generalization of factorial is defined by $[a]_\kappa^\lambda
= \prod_{(i,j)\in \kappa}(a + (j-1)/\lambda - (i-1))$.
Using the above notations, the density-density correlation function
at {\it finite} $N$ and $L$ is given by
\begin{equation}
\langle 0|\rho(x,t)\rho(0,0)|0\rangle =
{2\over \lambda^2}\sum_\kappa {|\kappa|^2 \over j_\kappa^\lambda}
{([0']_\kappa^\lambda)^2 [N]_\kappa^\lambda \over
[N+1/\lambda - 1]_\kappa^\lambda} e^{i2\pi|\kappa|x/L} e^{-itE_\kappa},
\label{finited}
\end{equation}
where $E_\kappa = \sum_{j=1}^N \kappa_j^2 +
\lambda\sum_{j=1}^N(N+1-2j)\kappa_j$ and
$j_\kappa^\lambda = \prod_{(i,j)\in \kappa}
(\kappa'_j-i+1+(\kappa_i-j)/\lambda)
(\kappa'_j-i+(\kappa_i-j+1)/\lambda)$.
The product in $[0']_\kappa^\lambda$ does not
include the pair $(i,j) = (1,1)$.  In deriving the
finite size correlation function, I use the following two
relations \cite{macdonald2,stand}
\begin{equation}
\sum_i z_i^n = {n\over \lambda} \sum_{|\kappa| = n} {[0']_\kappa^{\lambda}
\over j_\kappa^{\lambda}} J_\kappa^{1/\lambda}(\{z_i\}),
\label{thm1}
\end{equation}
\begin{equation}
\langle\kappa|\kappa'\rangle_{\lambda} =
j_\kappa^{\lambda}
{[N]_\kappa^{\lambda}\over [N+1/\lambda - 1]_\kappa^\lambda}
\delta_{\kappa,\kappa'}.
\label{norm}
\end{equation}
The first relation can be used to expand the density operator
$(1/L)\sum_{j=1}^N \delta(x - x_j)$ in terms of the Jack symmetric
polynomial.  The second relation gives
the normalization constants of all the
excited states parameterized by $\kappa$.
The DDDCF is now
easy to calculate since the eigenstate $\kappa$ evolves in time
only with a phase $\exp(-itE_\kappa)$.

For $\lambda = p/q$, the coefficient
$[0']_\kappa^\lambda$ is zero if
${\cal D}(\kappa)$ consists of more than $p$ columns
or more than $q$ rows.  Hence, in the thermodynamic limit a local density
operator acting on the ground state provokes only the minimal excitations
consisting of $q$ and $p$ quasi-particles and holes.

The large $N$ expansion of Eq.\ (\ref{finited})
can be carried out at fixed density $\rho = N/L$ and the leading
order term corresponding to the thermodynamic limit is given by
Eq.\ (\ref{ddc}).  In this limit,
a new super selection rule emerges and suppresses the states with
$|\kappa_i - \kappa_j| \approx O(1)$ to
$O(1/N^{2\lambda})$ or $O(1/N^{2/\lambda})$ depending on the value
of $\lambda$.  This means that the states with quasi-particles and
quasi-holes with same momenta (velocities) are suppressed.
There are some exotic exceptions to this rule. The details will
be published elsewhere.

The form of the ground state wave-function, Eq.\ (\ref{gs}), has lead
Haldane to suggest that while the apparent statistics can be modified
with a singular gauge transformation, the ``natural'' statistics of the SCM
are fractional, and that the particle excitations carry charge 1
and flux $\pi \lambda$ and the hole excitations charge $-1/\lambda$ and
flux $-\pi$ \cite{duncan2}.  Indeed, if a singular gauge transformation
\cite{laughlin} is applied, the ground state wave function can be rewritten as
\begin{equation}
\Psi_0 = \prod_{j>l}{(z_j-z_l)^\beta \over |z_j-z_l|^\beta}
|z_j-z_l|^{\lambda}
\prod_k z_k^{-\alpha(N-1)/2},
\label{singular}
\end{equation}
where the apparent statistical parameter is now $\theta = \pi \beta$.
Unlike in the two-dimensional case, the transformed wave
function remains as the ground state of the original Hamiltonian and so
are all the other eigenstates. The DDDCF is also unchanged.
The HPOGF, however, will be different for different choices
of statistics (see, for example, \cite{lenard}).
Hence, in order to
calculate the HPOGF, it will be necessary to
adapt Haldane's ``natural'' exchange statistics for the SCM particles.

In order to consider the fractional
exchange statistics, it is convenient to multiply
the wave function $\Psi$ by
an ``ordering function'' $\varphi(x_{P_1},\ldots,x_{P_N})$
which is just a book keeping
device for the phase factor that depends on the permutation
$P$ and is set to unity for the fundamental region
$x_1 < x_2 < \cdots < x_N$.  (In the case of Fermions the function is
just a product of Grassmann numbers.)  Since the ordering function
automatically keeps track of all the exchange phases,
the particle exchange operator $P_{ij}$ acting on the full wave function
amounts to simply exchanging the indices $i$ and $j$.  If the
``natural'' statistics is chosen, the phases arising from $\varphi$ and
$\Psi$ can always be set to cancel each other!

In the original SCM there is no physical process which allows particles
to exchange. (i.e. The wave function vanishes like
$|x_i - x_j|^\lambda$ as $x_i \rightarrow x_j$.)
A few years ago, Polychronakos \cite{poly} solved this problem by
introducing an analog of the Yangian generator \cite{hahal}
\begin{equation}
\pi_j = p_j + i\pi\lambda/L\sum_{i\ne j}\mbox{cot}(\pi(x_i-x_j)/L)P_{ij},
\label{pi}
\end{equation}
where $p_j$ is the ordinary momentum operator, and showed that
the Hamiltonian $H = \sum_j \pi_j^2$ is fully integrable and
is same as the SCM up to a modification $\lambda(\lambda - P_{ij})$
plus some trivial constant.  The new operator $\pi_j$
is the momentum operator for the pseudo-particles, and
the momenta $\sum_j k_j$ corresponds to the
eigenvalues of $\sum_j \pi_j$.
This new Hamiltonian should be considered as the model of
one-dimensional ``anyons'' with fractional exchange statistics.

The single particle destruction operation on the ground state of
$N+1$ ``anyons'' is, then, given by
\begin{equation}
\Psi(x)|0\rangle_{N+1} =
z^{-\lambda N/2}\prod_{j=1}^N
(z - z_j)^\lambda z_j^{-\lambda/2}|0\rangle_N,
\end{equation}
where $z = \exp(i2\pi x/L)$.  Now,
a similar technique used for the DDDCF can be employed to solve for
the HPOGF.  In this case, the contributing
partitions have no more than $p$ rows and $q-1$
columns (i.e. $p$ quasi-holes and $q-1$ quasi-particles).
Therefore, the ``natural'' exchange statistics of the
real particles is fully compatible with
the exclusion statistics of the elementary excitations.

In the thermodynamic limit the one-particle Green's function
(hole propagator) is given by
\begin{equation}
\langle 0| \Psi^\dagger(x,t) \Psi(0,0) |0\rangle =
\rho D e^{-i\pi\lambda \rho} \prod_{i=1}^{q-1}\left(\int_0^\infty dx_i\right)
\prod_{j=1}^p \left(\int_0^1 dy_j\right)
F(q-1,p,\lambda|\{x_i,y_j\}) e^{i(Q x-Et)},
\label{propagator}
\end{equation}
where $F(q-1,p,\lambda|\{x_i,y_j\})$ is given by Eq.\ (\ref{formfac})
and D by
\begin{equation}
D = {\lambda^{2p(q-1)} \Gamma(1+\lambda) \over
\lambda^{p-q+1} (q-1)!p!} A(q-1,p,\lambda).
\end{equation}
Q and E are same as before except for the number of $x_j$.
At integer values of $\lambda$ (i.e. $q=1$ case where only quasi-holes
are excited), based on the equal-time results of Forrester \cite{forrest}
Haldane made a conjecture \cite{duncan2} which agrees with this formula.
I conjecture that
the minimal form factor for any two-point correlation function
is given by $F(m,n,\lambda|\{x_i,y_j\})$ if the intermediate
states involve only $m$ quasi-particles
and $n$ quasi-holes.

In conclusion, the SCM is shown to possess the fractional {\it exclusion}
and {\it exchange} statistics.  The motifs representing the full spectrum
are constructed and used to demonstrate the exclusion statistics,
explicitly.  The fractional statistics in the SCM is also confirmed
by calculating the exact dynamical
density-density correlation function and one-particle Green's function
(hole propagator) at any rational interaction
coupling constant using the theory of Jack symmetric
orthogonal polynomial.  The details of the calculation including
the full Green's function will be published elsewhere.

While this Letter focuses on the fractional statistics aspect of
the SCM, the method for calculating the correlation functions developed
here could be of interest to a wide variety
of people working on the disordered electronic system, the quantum chaos,
the random matrix theory, 2D QCD, etc.

I thank F.D.M. Haldane, F. Wilczek, Y. M. Cho, T. Hwa, C. Johnson,
R. Narayanan, and R. Kamien for useful discussions.
This work is supported by DOE grant \#DE-FG02-90ER40542.

\end{document}